\documentclass[aps,prd,showpacs,groupedaddress,hyperref,amsmath,amssymb]{revtex4}

\usepackage{bm}% bold math 

\def\beq{\begin{equation}}
\def\eeq{\end{equation}}
\def\beqar{\begin{eqnarray}}
\def\eeqar{\end{eqnarray}}
\def\gp{\gamma^\mu p_\mu}
\def\ep{\mbox{\boldmath$\displaystyle\mathbf{\epsilon}$}}
\def\0{\mbox{\boldmath$\displaystyle\mathbb{O}$}}
\def\p{\mbox{\boldmath$\displaystyle\mathbf{p}$}}
\def\g{\mbox{\boldmath$\displaystyle\mathbf{g}$}}
\def\q{\mbox{\boldmath$\displaystyle\mathbf{q}$}}
\def\bv{\mbox{\boldmath$\displaystyle\mathbf{\varphi}$}}
\def\hp{\mbox{\boldmath$\displaystyle\mathbf{\widehat{\p}}$}}
\def\hv{\mbox{\boldmath$\displaystyle\mathbf{\widehat{\bv}}$}}
\def\z{\mbox{\boldmath$\displaystyle\mathbf{z}$}}
\def\hz{\mbox{\boldmath$\displaystyle\mathbf{\widehat{\z}}$}}
\def\s{\mbox{\boldmath$\displaystyle\mathbf{\sigma}$}}
\def\J{\mbox{\boldmath$\displaystyle\mathbf{J}$}}
\def\K{\mbox{\boldmath$\displaystyle\mathbf{K}$}}
\def\x{\mbox{\boldmath$\displaystyle\mathbf{x}$}}

\def\1{\mbox{\boldmath$\displaystyle\mathbb{I}$}}

\def\bls{\vspace{\baselineskip}}

\begin{document}
\title{Self-interacting Elko dark matter  with an axis of locality}

\author{D.~V.~Ahluwalia}
\email[]{dharamvir.ahluwalia@canterbury.ac.nz} 
\homepage[]{http://www2.phys.canterbury.ac.nz/editorial/}
\author{Cheng-Yang~Lee}
\email[]{cyl45@student.canterbury.ac.nz} 
\author{D.~Schritt} 
\email[]{dsc35@student.canterbury.ac.nz} 

\affiliation{Department of Physics and Astronomy,  Rutherford Building\\
University of Canterbury, 
 Private Bag 4800, 
Christchurch 8020, New
Zealand}

\date{6 October 2009}

\begin{abstract}
 This communication is a natural and nontrivial continuation of the 2005 work of Ahluwalia and Grumiller on Elko.  Here we report  that Elko breaks Lorentz symmetry in a rather subtle and unexpected way by containing a `hidden' preferred direction.  Along this preferred direction, a quantum field based on Elko enjoys locality. In the form reported here, Elko offers a mass dimension one fermionic dark matter with a quartic self-interaction and a preferred axis of locality. The locality result crucially depends on a judicious choice of phases. 
 \end{abstract}


\maketitle

\section{Introduction}
\label{Sec:Introduction}

The particle nature of dark matter is still unsettled. What we do
know is that it is expected to be endowed with a self
interaction~\cite{Spergel:1999mh,Wandelt:2000ad,Ahn:2004xt,Balberg:2002ue},
and that it defines, or couples to, an axis that has come to be known
as the axis of evil~\cite{Land:2006bn,Samal:2008nv,Frommert:2009qw}.
The latter aspect is most likely to be settled by Planck. The
indicated self-interaction would ordinarily suggest that dark matter
is some sort of scalar field. However, as shown
in~\cite{Ahluwalia:2004sz,Ahluwalia:2004ab}, the Elko quantum field
is endowed with mass-dimension one, a property that allows for
un-suppressed Elko self-interaction.  Further consequences of the mass
dimensionality of Elko are that its possible interactions with the
mass-dimension-three-half Dirac and Majorana fields are suppressed by
one order of unification scale and that it cannot enter the SM
doublets. This, along with the fact that Elko does not carry the
standard U(1) gauge invariance, renders Elko a natural dark matter
candidate~\cite{Ahluwalia:2004sz,Ahluwalia:2004ab}.

Here we report that Elko breaks Lorentz symmetry in a rather subtle
and unexpected way by containing a `hidden' preferred direction. All
inertial frames that move with a constant velocity along this
direction are physically equivalent. Along this direction, a quantum
field based on Elko enjoys locality.

Our discourse begins with a review of the SM matter fields in
Sec.~\ref{Sec:SM}. In section~\ref{Sec:MajoranaSpinors} we
recapitulate the known problems with the interpretation of Majorana
spinors as c-numbers, and argue that these problems evaporate under a
more careful examination~\cite{Ahluwalia:2004sz,Ahluwalia:2004ab}. The
pace is deliberately slow. The discussion is designed to provide the
right setting for the taken departure. Sections~\ref{Sec:Elko}
and~\ref{Sec:ElkoMDO} form the core of this communication. The
discussion on the Elko dual presented in Sec.~\ref{Sec:ElkoDual} is a
significant addition to the previous work on
Elko~\cite{Ahluwalia:2004sz,Ahluwalia:2004ab}. The dramatically
changed locality structure arises from certain phases and
identifications introduced in the Elko spinors at rest (see
Eqns.~\ref{eq:xi1}-\ref{eq:zeta2}). Section~\ref{Sec:ElkoKG} reminds
that Elko satisfies the Klein-Gordon, but not the Dirac, equation. The
Elko spin sums are given in Sec.~\ref{Sec:ElkoSpinSums}. These spins
sums are needed for examining the locality structure of the Elko
quantum fields and had to be re-evaluated due to the mentioned changes
in the Elko rest spinors.\footnote{This part of the result remains the
  same as that reported in
  Refs.~\cite{Ahluwalia:2004sz,Ahluwalia:2004ab}.} These carry the seeds
of the mentioned preferred direction. Section~\ref{Sec:ElkoMDO}
formally introduces the Elko quantum
fields. Section~\ref{Sec:ElkoIdentificationWithDM} makes an argument
to identify Elko with self-interacting dark matter that is endowed
with an axis of locality. In the form reported here, Elko offers a
dimension-one fermionic dark matter with self-interaction and a
preferred axis of locality. The locality result crucially depends on a
judicious choice of phases. The paper ends with summarising remarks
and questions in Sec.~\ref{Sec:ConcludingRemarks}. An appendix
provides supplementary information.

\subsection{The matter field underlying the SM}
\label{Sec:SM}

The matter field underlying the SM is a four-component spinor
field~\cite{Weinberg:1995mt} with historical origin in Dirac's
celebrated 1928 paper~\cite{Dirac:1928hu} \beq \Psi(x) =
\sum_\sigma\int \frac{d^3p}{(2 \pi)^{3}} \frac{1}{\sqrt{2
E(\p)}}\bigg[ \underbrace {u(x;\p,\sigma)}_{=u(\mathbf{p},\sigma)
e^{-i p^\mu x_\mu}} a(\p,\sigma) \;\;\;+ \underbrace
{v(x;\p,\sigma)}_{=v(\mathbf{p},\sigma) e^{+i p^\mu x_\mu}}
b^\dagger(\p,\sigma)\bigg] \eeq where $\sigma$ takes the values $\pm
1/2$. The zero-momentum coefficient functions may be symbolically
written as

  \beqar  u(0,1/2) &=& \left[\begin{array}{l} \uparrow \\
      \uparrow\end{array}\right],\hspace{.78cm} u(0,-1/2) =
  \left[\begin{array}{l} \downarrow \\ \downarrow\end{array}\right]
  \label{eq:u}\\  v(0,1/2) &=& \left[\begin{array}{l} \downarrow \\
      - \downarrow\end{array}\right],\hspace{.4cm} v(0,- 1/2) =
  \left[\begin{array}{l} - \uparrow \\ \uparrow\end{array}\right]
  \label{eq:v} \eeqar
  where 
  \beq \uparrow \stackrel{\rm def}{=} \sqrt{m}
  \left(\begin{array}{l}1\\0\end{array}\right),\quad
    \downarrow\stackrel{\rm def}{=} \sqrt{m} \left(\begin{array}{l} 0\\1
    \end{array}\right)\label{eq:updown}
    \eeq
    in the `polarisation basis'. In the helicity basis, these are
    eigenspinors of the helicity operator with a specific choice of
    phases. These phases are determined, e.g., by the locality
    condition.\footnote{In fact $\Psi(x)$ ceases to transform
      correctly under Poincar\'e transformations if these phases are
      chosen at random.}

    Without any reference to the Dirac equation (see
    Ref.~\cite{Weinberg:1995mt} for a detailed argument), the
    coefficient functions are determined from the condition that under
    the homogeneous Lorentz transformations the field components
    superimpose with other field components via spacetime-independent
    elements (of $4\times 4$ matrices). These matrices must furnish a
    finite dimensional representation of the homogeneous Lorentz
    group.

    The coefficient functions for arbitrary momentum are obtained by the
    action of the boost 
    \beq u(\p,\sigma) = \kappa\, u(0,\sigma)
    \label{eq:kappa} 
    \eeq
    where $\kappa:=\kappa_r\oplus\kappa_\ell$. The explicit expressions
    for $\kappa_r$ and $\kappa_\ell$ are given below.

    The only non-trivial freedom that $\Psi(x)$ still carries is the
    specialisation to the case where $ b^\dagger(\p,\sigma)$ is
    identified with $ a^\dagger(\p,\sigma)$. Otherwise, the Poincar\'e
    spacetime symmetries along with the symmetries of
    charge-conjugation, parity and time-reversal and the demand of
    locality uniquely determine the field $\Psi(x)$. Seen in this light
    the field coefficients $u(\p,\sigma)$ and $v(\p,\sigma)$ are
    eigenspinors of the $\gamma^\mu p_\mu$ operator with eigenvalues
    $+m$ and $-m$ respectively.

    The annihilation of the field $\Psi(x)$ by the Dirac operator
    $(i\gamma^\mu \partial_\mu - m \1)$ follows as a result of this
    structure. The Dirac equation is not assumed. Rather, it emerges as
    a direct consequence of the merger of quantum mechanics and
    Poincar\'e spacetime symmetries for spin one half. The apparent
    simplicity of the Dirac field can be somewhat misleading to the
    uninitiated. For instance, a change in sign in the right-hand side
    of the expression for $v(0,- 1/2)$ in Eqn.~(\ref{eq:v}) yields a
    quantum field that is nonlocal when $b^\dagger(\p,\sigma)$ is
    identified with $a^\dagger(\p,\sigma)$. Even though the mentioned
    change in phase does not destroy the locality in the original field,
    it does violate spacetime symmetries in a hidden way. A systematic
    study of such subtle loss of symmetries and locality remains largely
    unexplored.

    For historical reasons the field $\Psi(x)$ is known as the Dirac
    field, while the identification of $b^\dagger(\p,\sigma)$ with
    $a^\dagger(\p,\sigma)$ yields what has come to be known as the
    Majorana field~\cite{Dirac:1928hu,Majorana:1937vz}. The coefficient
    functions $u(\p,\sigma)$ and $v(\p,\sigma)$ are the usual Dirac
    spinors. They can be interpreted as being a direct sum of the r-type
    and $\ell$-type Weyl spinors with specific helicities and phases.

    \subsection{Majorana spinors: A critique}
    \label{Sec:MajoranaSpinors}

    History clearly demarcates the introduction of the Majorana
    field. It was introduced in 1937 under the pen of Ettore Majorana~
    \cite{Majorana:1937vz}. As regards Majorana spinors, we (i.e, the
    authors) do not know of their historical birth.

    While in the operator formalism of quantum field theory Dirac
    spinors can be treated as c-numbers, it is curious that Majorana
    spinors must be treated as G-numbers. This is deemed necessary, due
    to what are considered otherwise unavoidable problems (consider for
    instance Aitchison and Hay's attempt to construct a Hamiltonian
    density~\cite{Aitchison:2004cs}). What further adds to the problem
    is that taken by itself a Majorana spinor is nothing but a Weyl
    spinor in the four-component form. As shown by Ahluwalia and
    Grumiller~\cite{Ahluwalia:2004sz,Ahluwalia:2004ab} both of these
    problems can be circumvented. The solution to the first problem is
    found by acting the Dirac operator on a Majorana spinor and finding
    that, apart from a numerical factor, this operation yields another
    Majorana spinor (but not the same as the original one). The action
    of the Dirac operator again returns back the original Majorana
    spinor. This suggests that the problem is not with the Majorana
    spinors but with the Lagrangian density Aitchison and Hey
    assumed~\cite{Aitchison:2004cs}! The latter of the two mentioned
    problems also has a similar solution. The usual set of Majorana
    spinors are a set of two spinors, and both of these have eigenvalue
    of unity under the operation of charge conjugation operator. This is
    the self-conjugate set. However, as pointed out in
    Ref.~\cite{Ahluwalia:2004sz,Ahluwalia:2004ab} there also exists the
    anti self-conjugate set. Once that is done the complete set of four
    spinors -- the Elko (for \textbf{E}igenspinoren des
    \textbf{L}adungs\textbf{k}onjugations\textbf{o}perators) -- span the
    four-dimensional representation space of spin one half and come to
    par with the Dirac spinors.

    Thus we note again that while in the operator formalism of quantum
    field theory Dirac spinors can be treated as c-numbers, it is
    curious that Majorana spinors must be treated as G-numbers but now
    ask if this distinction hides something fundamental about Majorana
    spinors, and what the Grassmann aspect has to do with their
    fermionicness. After all the operator formalism lends itself much
    more directly to the quantum mechanical description and fermionic
    aspect of Dirac particles, using Dirac spinors, is implemented
    through the anticommutation relations associated with the creation
    and destruction operators. Why is that Majorana spinors, without
    giving them the Grassmann character, have not found their place in a
    quantum field operator. In an attempt to answer this question we are
    led to a better understanding of the canonical wisdom, and this
    allows us the necessary departure in which Majorana spinors as
    c-numbers play an important role.

    \bls 

    We now move towards greater preciseness by rephrasing some of the
    canonical wisdom and by examining it in greater detail. As we
    proceed through the critique we shall find that the mentioned
    obstacles can be surmounted. We do not refrain from making explicit
    the cost at which this happens. Whether or not one ought to accept
    the cost, in part, is a matter of experiments to decide. At the very
    least we shall know as to what it is that we reject if we choose to
    confine to the canonical wisdom alone.

    In the received wisdom, the Majorana spinors arise as follows. If
    $\phi_\ell$ is a massive Weyl spinor of $\ell$-type, then $\sigma_2
    \phi_\ell^\ast$ transforms as a r-type Weyl spinor. For this
    reason~\cite[p.20]{Ramond:1981pw}, we can construct a special type
    of four-component spinor called a Majorana spinor
    \begin{equation} \psi_M = \left(\begin{array}{cc} -\sigma_2
	\phi_\ell^\ast \\ \phi_\ell\end{array}\right)
	\label{eq:MajoranaSpinor} \end{equation} 
    It is self-conjugate under charge conjugation. For $\phi_\ell$ there
    are two choices, a positive helicity and a negative helicity. As
    such, we have two rather than four four-component spinors. Thus the
    folklore: the Majorana spinor is a Weyl spinor in four component
    form~\cite{Ramond:1981pw}. It is self evident, and remains
    unquestioned in our discourse.

    An immediate sign of trouble appears if one naively introduces a
    Lagrangian density $\mathcal{L}_M =
    \overline{\psi}_M\left(i\gamma^\mu \partial_\mu - m
    \right)\psi_M$. The usual route at this stage is to treat the Weyl
    spinorial components as Grassmann numbers, otherwise one encounters
    the often quoted problems~\cite[App. P]{Aitchison:2004cs}. The
    Ahluwalia-Grumiller work in
    references~\cite{Ahluwalia:2004sz,Ahluwalia:2004ab} strongly
    indicates that this approach may be hiding certain fundamental
    properties of Majorana spinors. Or, to put it more precisely, having
    gone the Grassmannian route we may have escaped a rich and fertile
    ground where Majorana spinors are treated as c-number spinors. To
    unearth these aspects here we will treat the massive Weyl spinors as
    2-component eigenspinors of the helicity
    operator~\cite[p. 111]{ORaifeartaigh:1977lo} and the fermionic
    statistics shall be implemented through the canonical field operatic
    formalism~\cite{Weinberg:1995mt,Ryder:1985wq}, and not by treating
    them as Grassmann fields.\footnote{The Grassmann field is not to be
      confused with a quantum field of the operator formalism.  The latter
      carries a precise meaning as explained in detail by
      Weinberg~\cite[Ch. 5]{Weinberg:1995mt}. This distinction, however
      (to a reader's confusion), seems to disappear in the closing volume
      of the Weinberg's triology~\cite[pp. 58-59]{Weinberg:2000cr}.}  The
    Elko formalism was born in this spirit and it attended to a
    widespread, but rarely spoken, discontent with abandoning Majorana
    spinors as c-numbers as the first author realised while at Los
    Alamos.

    A straightforward calculation now shows that, (i) under the Dirac
    dual the norm $\overline\psi_M\psi_M$ identically vanishes (so, no
    Dirac mass term); and (ii) in the momentum space, $\psi_M$ is not an
    eigenspinor of the $\gamma_\mu p^\mu$ operator: $\gamma_\mu p^\mu
    \psi_M \ne \pm m \psi_M$ (so, Majorana spinors do not satisfy Dirac
    equation~\cite[App. P]{Aitchison:2004cs}. Combined, this already
    hints that securing a mass dimension three half fermionic field
    expanded in terms of Majorana spinors may not be
    possible.\footnote{The question that this exercise cannot be
      implemented with self conjugate spinors alone shall be addressed
      below.}  The lesson to be learned eventually is that it is
    \textit{not} sufficient that one considers the ``simplest candidates
    for a kinematic spinor term'' (or, for that matter any field), as
    found in almost\footnote{An exception being Weinberg's triology on
      the theory of quantum
      fields~\cite{Weinberg:1995mt,Weinberg:1996kr,Weinberg:2000cr}.}
    every text book on quantum field theory\footnote{The quote here is
      from Ramond's primer.} but that the associated Green function be
    proportional to the vacuum expectation value of (the time ordered
    product) of certain fields operators. This lesson, we think, has a
    much larger significance in that Lagrangian densities must be
    derived and not assumed. Neglecting this may induce all sort of
    pathologies. How this task is to be accomplished~\textemdash ~at
    least for spin one half~ \textemdash~ is one the wider contributions
    of this communication.

    The assertion about reduction in the degrees of freedom for Majorana
    spinors also faces trouble if one notes that the relevant charge
    conjugation operator has not one, but two, real eigenvalues, $+ 1$
    (giving the usual self-conjugate Majorana spinors), and $-1$. There
    is no physical or mathematical reason to abandon, or project out,
    the latter. The sense in which the folklore still survives is that
    by an appropriate similarity transformation half of these (i.e.,
    those corresponding to the positive eigenvalue) can be morphed into
    real 4-component spinors, while those corresponding to the negative
    eigenvalue can be transformed into pure imaginary 4-component
    spinors.

    \section{Elko: The departure from Grassmann interpretation of
      Majorana spinors}
    \label{Sec:Elko}

    The Grassmann interpretation of the Majorana spinors is elegant. It
    is mathematically sound. It has found wide spread applications in
    modern quantum field theory. Yet, it breaks with the tradition of
    field operatic formalism which would have required these spinors to
    be c-number coefficient functions in a field. To implement this
    programme Ahluwalia and Grumiller undertook a new effort in
    references~\cite{Ahluwalia:2004sz,Ahluwalia:2004ab}.  They
    introduced a complete set of dual helicity eigenspinors of the
    charge conjugation operator for spin one half. In their formalism
    there are four, rather than two, 4-components spinors. This is the
    first point of departure.  For the mentioned set of spinors they
    introduced the name Elko.  The new name, as already mentioned, was
    taken from German and stood for \textbf{E}igenspinoren des
    \textbf{L}adungs\textbf{k}onjugations\textbf{o}perators. It was
    necessitated to mark the distinct physical and mathematical content
    of the introduced departure; and to avoid confusion with the
    literature on Majorana spinors and fields where Grassmannian
    interpretation reigns.

    Grassmannisation of Majorana spinors is a deep and conceptually
    nontrivial element of theoretical landscape.  The quantum-mechanical
    field it introduces is not a quantum field in the sense of Weinberg
    (specifically, in the sense of Weinberg's
    monograph~\cite{Weinberg:1995mt}).  At the same time the uniqueness
    of Dirac field, modulo its specilisation to the Majorana field, also
    implies that the programme we embark upon shall necessarily contain
    an element that breaks Lorentz invariance in some way.  This feature
    had remained hidden in our previous discourse.  We now explicitly
    unearth it. It has the potential to open up an entirely new
    perspective on dark matter, and its physics. The decision being in
    the hands of experiments. To a pure theoretician the interest might
    be in its mathematical structure.

    In this communication we confine our primary attention to spin one
    half, but we construct Elko in such a way that the procedure
    immediately generalises to all spins. This is facilitated by the use
    of Wigner's time reversal operator $\Theta$, rather than the Pauli
    matrix $\sigma_2$ that appears in Ramond's primer in the context of
    Majorana spinors. We shall use the phrase Elko for spinors as well
    as the quantum fields constructed from them.  The context shall be
    assumed to remove any ambiguity.

    \subsection{Construction of Elko}

    To construct Elko it is first necessary to introduce the charge
    conjugation operator. This we do as follows.  Under parity, $P$, $\x
    \to \x^\prime = - \x$, $\bv \to -\bv$ and $\s \to \s$. Consequently,
    an examination of Eqs.~(\ref{eq:boostL}) and (\ref{eq:boostR}),
    yields $\kappa_\ell \stackrel{P}{\leftrightarrow} \kappa_r$.  This
    observation suffices to give the action of parity on the
    $r\oplus\ell$ representation space up to a phase
    \begin{equation}
      S(P)= \exp[i \vartheta] \underbrace{\left(\begin{array}{cc}
	  \mathbb{O} & \1 \\ \1 & \mathbb{O}
	\end{array}\right)}_{\gamma^0}\mathcal{R},\quad \vartheta\in
      \mathbb{R}
    \end{equation}
    With $\p:= p\left(\sin(\theta)\cos(\phi),
    \sin(\theta)\sin(\phi),\cos(\theta)\right)$, the
    $\mathcal{R}=\{\theta\to\pi-\theta,\phi\to\pi+\phi,p\to p\}$.  If
    care is taken that the eigenvalues of the helicity operator change
    sign under $P$, the arguments given in Ref.~\cite{Ahluwalia:2004ab}
    fix the phase $\exp[i\vartheta]$ to be $i$. The $S(P)$ now has four
    doubly-degenerate eigenspinors, carrying opposite eigenvalues of
    $S(P)$ \textemdash~call these $u$ and $v$ sectors.  The operator
    \begin{equation}
      \mathcal{C}=\left(\begin{array}{cc} \mathbb{O} & i\Theta \\
	-i\Theta & \mathbb{O}\end{array}\right) K\label{eq:Wigner}
    \end{equation}
    where $K$ is the complex conjugation operator, formally interchanges
    the opposite parity sectors: $u
    \stackrel{\mathcal{C}}{\leftrightarrow} v$. It is apparent that
    $\mathcal{C}$ is the standard charge conjugation operator of Dirac. In the context of Eqn.~(\ref{eq:Wigner}) Wigner's time
      reversal operator $\Theta$ is defined as $\Theta \J \Theta^{-1} =
      - \J^\ast $ where $\J$ are a set of rotation generators for the
      representation space under consideration. For spin one half,
      $\Theta\left[\s/2\right]\Theta^{-1} = - \left[\s/2\right]^\ast$.
      We use the realisation
      \[
      \Theta = \left(
      \begin{array}{lr}
	0 & -1 \\ 1 & 0
      \end{array}
      \right)
      \]

    To construct Elko, let $\phi_\ell(\p)$ be a $\ell$-type Weyl spinor
    of spin one half.  Under a Lorentz boost, it transforms as
    $\phi_\ell(\p) = \kappa_\ell \phi_\ell(\ep)$ with
    \begin{equation} 
      \kappa_\ell= \exp\left(-\frac{\s}{2}\cdot\bv\right) = \sqrt{
	\frac{E+m}{2 m}}\left(\1 - \frac{\s\cdot\p}{E + m}\right)
      \label{eq:boostL}
    \end{equation}
    The $\ep$ is defined as $\p\vert_ {p\to 0}$, and not as $\p\vert_
    {p=0}$. This restriction can be removed, if necessary (for example,
    by working in `polarisation basis' which then comes with its own
    subtleties).  The basic results remain unaltered. In the usual
    notation, the boost parameter $\bv$ is defined as
    \beq \cosh\varphi = \frac{E}{m} = \gamma =
    \frac{1}{\sqrt{1-\beta^2}},\quad \sinh\varphi = \frac{p}{m} =
    \gamma\beta,\quad\hv=\hp \eeq
    By $\s=(\sigma_1,\sigma_2,\sigma_3)$ we denote the Pauli
    matrices. The symbol $\1$ represents an identity matrix, while in
    what follows $\mathbb{O}$ shall be used for a null matrix (their
    dimensionality shall be apparent from the context).  For
    $\phi_\ell(\p)$ we have two possibilities: $\s\cdot\hp
    \,\phi_\ell^\pm(\p) = \pm\, \phi_\ell^\pm(\p)$.

    Following Ref.~\cite{Ahluwalia:2004ab} we now note that under a
    Lorentz boost, $\vartheta \Theta \phi_\ell^\ast(\p)$ transforms as a
    r-type Weyl spinor, $\left[\vartheta \Theta
      \phi_\ell^\ast(\p)\right] = \kappa_r \left[\vartheta \Theta
      \phi_\ell^\ast(\ep)\right]$, with
    \begin{equation} 
      \kappa_r = \exp\left(+\frac{\s}{2}\cdot\bv\right) = \sqrt{
	\frac{E+m}{2 m}}\left(\1 + \frac{\s\cdot\p}{E + m}\right)
      \label{eq:boostR}
    \end{equation}
    where $\vartheta$ is an unspecified phase to be determined below.
    The helicity of $\vartheta \Theta \phi_\ell^\ast(\p)$ is {\em
      opposite} to that of $\phi_\ell(\p)$,
    \begin{equation} \s\cdot\hp \left[ \vartheta \Theta
	\left(\phi_\ell^\pm(\p)\right)^\ast \right] = \mp \, \left[
	\vartheta \Theta \left(\phi_\ell^\pm(\p)\right)^\ast \right]
    \end{equation}

    The argument that led to \emph{two} Majorana spinors, now instead
    takes us to their cousins, the \textit{four} 4-component spinors
    with the general form
    \begin{equation}
      \chi(\p)= \left(\begin{array}{c} \vartheta \Theta
	\phi_\ell^\ast(\p)\\ \phi_\ell(\p)
      \end{array}\right)\label{eq:taup}
    \end{equation}
    The $\chi(\p)$ become eigenspinors of the charge conjugation
    operator, Elko, with real eigenvalues if the phase $\vartheta$ is
    restricted to $\pm \, i$
    \begin{equation}
      \mathcal{C}\; \chi(\p)\Big\vert_{\vartheta=\pm i} = \pm
      \chi(\p)\Big\vert_{\vartheta=\pm i} \label{eq:tau}
    \end{equation}

    \vspace{0.89cm} 

    One can motivate the well-known Dirac spinors in a parallel manner;
    as eigenspinors of the parity operator $S(P)$. In that case the
    right- and left-transforming components are necessarily endowed with
    the same helicity.  For Elko, the right- and left- transforming
    components carry opposite helicity. So, whereas Dirac spinors may
    exist as eigenspinors of the helicity operator, the Elko
    cannot. This eventually reflects in many results that we arrive at.

    To give Elko a concrete form, we adopt the global phases so that `at
    rest' the $\ell$-type Weyl spinors take the form\footnote{In a
      separate calculation we have taken the rest spinors
      $\phi_\ell^\pm(\ep)$ to be eigenspinors of the operator
      $(\s/2)\cdot\hz$ and confirmed the results reported here. In that
      event the $\phi_\ell$ must carry certain global phases.}
    \begin{subequations}
      \begin{eqnarray}
	\phi_\ell^+(\ep)&=& \sqrt{m}\left(\begin{array}{c}
	  \cos(\theta/2) e^{-i \phi/2}\\ \sin(\theta/2) e^{i \phi/2}
	\end{array}\right)\label{eq:phiplus}\\
	\phi_\ell^-(\ep) &=& \sqrt{m}\left(\begin{array}{c}
	  -\sin(\theta/2) e^{-i \phi/2}\\ \cos(\theta/2) e^{i \phi/2}
	\end{array}\right) \label{eq:phiminus}
      \end{eqnarray}
    \end{subequations}
    Equations~(\ref{eq:phiplus}-\ref{eq:phiminus}), when coupled with
    Eqn.~(\ref{eq:taup}), allow us to explicitly introduce the
    self-conjugate spinors ($\vartheta=+i$) and anti self-conjugate
    spinors ($\vartheta=-i$) at rest\footnote{ If one wishes, one could
      replace the $\pm$ signs (after the symbol $:=$) that appear in
      Eqns. (\ref{eq:xi1}-\ref{eq:zeta2}) by four different phases of the
      form $\exp(i \vartheta_{\{\mp,\pm\}})$. The latter are immediately
      reduced to $+1$ or $-1$, otherwise the defining-requirement
      (\ref{eq:tau}) is violated. The rest of the freedom, i.e. whether to
      chose $+$ sign or the $-$ sign, can then be constrained later by the
      demand of locality.  This observation is as true for Elko fields
      that we construct as it is for the counterparts of
      Eqns. (\ref{eq:xi1}-\ref{eq:zeta2}) in the the construction of the
      Dirac and Majorana fields. The precise condition that fixes these
      phases in Weinberg's analysis is the transformation properties of
      the fields under the Poincar\'e
      group~\cite{Weinberg:1995mt}. Locality and spacetime symmetries of
      the fields are deeply intertwined.}
    \begin{subequations} 
      \begin{eqnarray} \xi_{\{-,+\}}(\ep) &:=&
	+\; \chi(\ep)
	\big\vert_{\phi_\ell({\displaystyle\mathbf{\epsilon}})
	  \to\phi_\ell^+({\displaystyle\mathbf{\epsilon}}),\;\vartheta=+i}
	\label{eq:xi1}
	\\ \xi_{\{+,-\}}(\ep) &:=& +\; \chi(\ep)
	\big\vert_{\phi_\ell({\displaystyle\mathbf{\epsilon}})
	  \to\phi_\ell^-({\displaystyle\mathbf{\epsilon}}),\;\vartheta=+i}
	\label{eq:xi2}	
	\\ \zeta_{\{-,+\}}(\ep) &:=& +
	\;\chi(\ep)\big\vert_{\phi_\ell({\displaystyle\mathbf{\epsilon}})
	  \to\phi_\ell^-({\displaystyle\mathbf{\epsilon}}),\;\vartheta=-i}
	\label{eq:zeta1}
	\\ \zeta_{\{+,-\}}(\ep) &:=& -\;
	\chi(\ep)\big\vert_{\phi_\ell({\displaystyle\mathbf{\epsilon}})
	  \to\phi_\ell^+({\displaystyle\mathbf{\epsilon}}),\;\vartheta=-i}
	\label{eq:zeta2} 
      \end{eqnarray} 
    \end{subequations}
    For comparison with equations (\ref{eq:u}-\ref{eq:updown}), the
    above in `polarisation basis' may be written as 
    \beqar &&
    \xi_{\{-,+\}}(\ep) = \left[\begin{array}{c} i \Downarrow \\ \Uparrow
      \end{array}\right],\quad
    \xi_{\{+,-\}}(\ep) = \left[\begin{array}{c} - i \Uparrow \\
	\Downarrow
      \end{array}\right] \label{eq:xi}\\
    && \zeta_{\{-,+\}}(\ep) = \left[\begin{array}{c} i \Uparrow \\
	\Downarrow
      \end{array}\right],\quad
    \zeta_{\{+,-\}}(\ep) = - \left[\begin{array}{c} - i \Downarrow \\
	\Uparrow
      \end{array}\right] \label{eq:zeta}
    \eeqar 
    The $\Uparrow$ and $\Downarrow$ differ from $\uparrow$ and
    $\downarrow$ of Eqn.~(\ref{eq:updown}) by certain phases (otherwise,
    as is appropriate for the `polarisation basis,' they are identical
    when $\theta$ and $\phi$ are set to zero).  In the context of
    Weinberg's work on the uniqueness of Dirac field (modulo its
    `Majoranaisation' in the sense of the 1937 original paper of
    Majorana~\cite{Majorana:1937vz}), a comparison with
    Eqns.~(\ref{eq:u}-\ref{eq:v}) already tells us that a quantum field
    that fully respects Lorentz symmetries cannot be built in terms of
    $\xi$ and $\zeta$ Elko spinors. The task then is to unearth this
    violation, and see how strong, or how weak, the said violation is.

    The $\xi(\p)$ and $\zeta(\p)$ for an arbitrary momentum are now
    readily obtained
    \begin{equation}
      \xi(\p) = \kappa\, \xi(\ep),\; \zeta(\p) = \kappa\,
      \zeta(\ep),\quad \kappa:=\kappa_r\oplus\kappa_\ell
    \end{equation}

    \subsection{A systematic construction of Elko dual, orthonormality, and completeness}
    \label{Sec:ElkoDual}

    The norm of Elko under the Dirac dual $\overline{\chi}(\p):=
    [\chi(\p)]^\dagger \gamma^0$ identically vanishes. However, it is
    more appropriate to seek a `metric' $\eta$ such that the product $
    [\chi_\imath (\p) ]^\dagger \eta \chi_\jmath(\p) $ \textemdash~with
    $\chi_\imath(p)$ as any one of the four Elko spinors
    \textemdash~remains invariant under an arbitrary Lorentz
    transformation. This requirement can be readily shown to translate
    into the following constraints on $\eta$
    \begin{equation} 
      \left[J_i, \eta\right] =0, \quad \left\{K_i,\eta\right\} =0
    \end{equation}
    Since the only property of the generators of rotations and boosts
    that enters the derivation of the above constraints is that
    $\J^\dagger = \J$ and $\K^\dagger = - \K$, the result applies to all
    \emph{finite} dimensional representations of the Lorentz group. It
    need not be restricted to Elko alone. Seen in this light, there is
    no non-trivial solution for $\eta$ either for the r-type or the
    $\ell$-type Weyl spinors.  For $r\oplus\ell$ representation space,
    the most general solution is found to carry the form
    \begin{equation}
      \eta = \left[\begin{array}{cccc} 0 & 0 & a & 0 \\ 0 & 0 & 0 & a \\
	  b & 0 & 0 & 0\\ 0 & b & 0 & 0
	\end{array}\right]  \label{eq:metric}
    \end{equation}
    It is now convenient to introduce the notation $\chi_1(\p) :=
    \xi_{\{-,+\}}(\p)$, $\chi_2(\p) := \xi_{\{+,-\}}(\p)$, $\chi_3(\p)
    := \zeta_{\{-,+\}}(\p)$, and $\chi_4(\p) :=
    \zeta_{\{+,-\}}(\p)$. Then sixteen values of $[\chi_\imath (\p)
    ]^\dagger \eta \chi_\jmath(\p)$ as $\imath$ and $\jmath$ vary from 1
    to 4 are given in Table 1.
    \begin{table}[htdp]
      \caption{The values of $[\chi_\imath (\p) ]^\dagger \eta
	\chi_\jmath(\p)$ evaluated using $\eta$.  The $\imath$ runs from
	1 to 4 along the rows and $\jmath$ does the same across the
	columns.}
      \begin{center}
	\begin{tabular}{|c|c|c|c|}
	  \hline 0 & $- i m (a + b)$ & $- i m(a - b)$ & 0 \\ \hline $i m
	  (a + b)$ & 0 & 0 & $- i m (a - b)$\\ \hline $- i m (a - b) $&
	  0 & 0 &$ i m (a + b)$\\ \hline 0 & $- i m (a - b) $& $- i m(a
	  + b)$ & 0 \\ \hline
	\end{tabular}
      \end{center}
      \label{dual}
    \end{table}
  
    To allow for the possibility of parity covariance we set $b=a$ (this
    treats $r$ and $\ell$ Weyl spaces on the same footing).  To make the
    invariant norms real, we give $a$ and $b$ the common value of $\pm
    i$; resulting in $\eta = \pm i\gamma^0$. In what follows the choice
    of the signs shall be dictated by the convenience of book keeping.

    \bls

    Guided by these results we now introduce the \emph{Elko dual}
    \begin{equation}
      {{\stackrel{\neg}{\chi}}_{\{\mp,\pm\}}(\p)} := \mp i
      \left[\chi_{\{\pm,\mp\}}(\p)\right]^\dagger \gamma^0
      \label{eq:dualchi} \\
    \end{equation}
    Under the new dual the orthonormality relations read
    \begin{subequations}
      \begin{eqnarray}
	&& {{\stackrel{\neg}{\xi}}_{\alpha}(\p)}\,
	\xi_{\alpha^\prime}(\p) = + \,2 m \delta_{\alpha\alpha^\prime}
	\label{eq:xinorm} \\
	&& {{\stackrel{\neg}{\zeta}}_{\alpha}(\p)} \,
	\zeta_{\alpha^\prime}(\p) = -\, 2 m \delta_{\alpha\alpha^\prime}
	\label{eq:zetanorm}
      \end{eqnarray}
    \end{subequations}
    along with ${{\stackrel{\neg}{\xi}}_{\alpha}(\p)}\,
    \zeta_{\alpha^\prime}(\p) = 0$, and
    ${{\stackrel{\neg}{\zeta}}_{\alpha}(\p)}\, \xi_{\alpha^\prime}(\p) =
    0 $.  The dual helicity index $\alpha$ ranges over the two
    possibilities: $\{+,-\}$ and $\{-,+\}$, and $-\{\pm,\mp\}:=
    \{\mp,\pm\}$. The completeness relation
    \begin{equation}
      \frac{1}{2 m}\sum_\alpha\left[ \xi_{\alpha}(\p)\,
	{{\stackrel{\neg}{\xi}}_{\alpha}(\p)}- \zeta_{\alpha}(\p)\,
	{{\stackrel{\neg}{\zeta}}_{\alpha}(\p)}\right] =
      \1\label{eq:completeness}
    \end{equation}
    establishes that we need to use {\em both} the self-conjugate as
    well as the anti self-conjugate spinors to fully capture the
    relevant degrees of freedom.

    \subsection{Elko satisfy the Klein-Gordon, not Dirac, equation}
    \label{Sec:ElkoKG}

    Because we are going to encounter several unexpected results, we
    pause to examine the behaviour of $\xi(\p)$ and $\zeta(\p)$ spinors
    under the action of the operator $\gamma^\mu p_\mu$. This brute
    force exercise serves the pedagogic purpose of countering some of
    prejudices that some readers may inevitably carry from their prior
    studies. Additionally, in the context of Aitchison and Hey's concern
    that one encounters a problem in constructing a Lagrangian density
    for Majorana spinors if they are not treated as Grassmann
    variables~\cite[App. P]{Aitchison:2004cs} , we provide the origin of
    that concern and offer a solution.

    We already have explicit expressions for $\xi(\p)$ and $\zeta(\p)$
    spinors. On these we act $\gamma^\mu p_\mu$ and find the following
    identities
    \begin{subequations}
      \begin{eqnarray}
	&&\gp \xi_{\{-,+\}}(\p) = +i m \xi_{\{+,-\}}(\p)\quad
	\Leftrightarrow \quad \gp \chi_1(\p) = +i m \chi_2(\p)
	\label{eq:a} \\ && \gp \xi_{\{+,-\}}(\p) = - i m
	\xi_{\{-,+\}}(\p) \quad \Leftrightarrow \quad \gp \chi_2(\p) = -
	i m \chi_1(\p) \label{eq:b} \\ &&\gp \zeta_{\{-,+\}}(\p) = - i m
	\zeta_{\{+,-\}}(\p) \quad \Leftrightarrow \quad \gp \chi_3(\p) =
	- i m \chi_4(\p)
	\label{eq:c} \\
	&& \gp \zeta_{\{+,-\}}(\p) = +i m \zeta_{\{-,+\}}(\p)\quad
	\Leftrightarrow \quad \gp \chi_4(\p) = +i m \chi_3(\p)
	\label{eq:d}
      \end{eqnarray}
    \end{subequations}
    Applying $\gamma^\nu p_\nu$ to Eqn. (\ref{eq:a}) from the left and
    then using (\ref{eq:b}) on the resulting right hand side, and
    repeating the same procedure for the remaining equations we get
    \begin{equation}
      \left(\gamma^\nu\gamma^\mu p_\nu p_\mu - m^2\right)
      \xi_{\{\mp,\pm\}}(\p) = 0,\quad \left(\gamma^\nu\gamma^\mu p_\nu
      p_\mu - m^2\right) \zeta_{\{\mp,\pm\}}(\p) = 0.
    \end{equation}
    Now using $\{\gamma^\mu,\gamma^\nu\} = 2 \eta^{\mu\nu}$, yields the
    Klein-Gordon equation (in momentum space) for the $\xi(\p)$ and
    $\zeta(\p)$ spinors.  Aitchison and Hey's concern is thus
    overcome. The problem resides in the approach of constructing the
    ``simplest candidates for a kinematic spinor term.''

    \subsection{Elko spin sums: a preferred axis}
    \label{Sec:ElkoSpinSums}

    We now look at the spin sums in Eqn.~(\ref{eq:completeness})
    separately.  These evaluate to
    \begin{subequations} \begin{eqnarray} && \sum_\alpha
	\xi_\alpha(\p) \stackrel{\neg}{\xi}_\alpha(\p) = m \left[
	  {\mathcal G}(\p) + \1 )\right]
	\label{eq:spinsumxi}\\ 
	&& \sum_\alpha \zeta_\alpha(\p)
	\stackrel{\neg}{\zeta}_\alpha(\p) = m \left[{\mathcal G}(\p)
	  -\1\right]\label{eq:spinsumzeta} \end{eqnarray}
    \end{subequations}
    which together {\em define} ${\mathcal G}(\p)$.  A \textit{direct
      evaluation} of the left hand side of the above equations gives
    \begin{equation} {\mathcal G}(\p)= i \left( \begin{array} {cccc} 0 &
	0 & 0 & - e^{-i \phi}\\ 0 & 0 & e^{i\phi} & 0 \\ 0 & - e^{-i
	  \phi}&0 & 0\\ e^{i\phi} & 0 & 0 & 0 \end{array} \right)
      \label{eq:gp}
    \end{equation} 
    For later reference, we note that ${\mathcal G}(\p)$ is an odd
    function of $\p$
    \begin{equation}
      {\mathcal G}(\p) = -\, {\mathcal G}(-\p)
    \end{equation} 
    But since ${\mathcal G}(\p)$ is independent of $p$ and $\theta$, it
    is more instructive to translate the above expression into
    \begin{equation} {\mathcal G}(\phi) = -\, {\mathcal
	G}(\pi+\phi) \label{eq:Gzimpok}
    \end{equation} 

    This serves to define a preferred axis, $z_e$.\footnote{The
      accompanying $x_e$ and $y_e$ axis help to define a preferred
      frame.}  Another hint for a preferred axis arises when one notes
    that the Elko spinorial structure does not enjoy covariance under
    usual local $U(1)$ transformation with phase $\exp(i
    \alpha(x))$. However, $U_E(1) = \exp(i\gamma^0\alpha(x))$
    ~\textemdash~and not $ U_M(1) = \exp(i\gamma^5\alpha(x))$ as one
    would have thought~\cite[p. 72]{Marshak:1969re}~\textemdash~
    preserves various aspects of the Elko structure. Similar comments
    apply to the non-Abelian gauge transformations of the SM.

    For a comparison with the Dirac counterpart
    (App. \ref{App:MisleadingDerivation}), we define $g^\mu:=(0,\g)$
    with $\g = - [1/\sin(\theta)] \partial \hp/\partial\phi =
    (\sin\phi,-\cos\phi,0)$. Note may be taken that $g^\mu$ is a unit
    spacelike four-vector, $g_\mu g^\mu = -1$.  Furthermore, $g_\mu
    p^\mu=0$.  In terms of $g^\mu$, $\mathcal{G}(\p)$ may be written as
    \begin{equation} {\mathcal G}(\p) =
      \gamma^5(\gamma_1\sin\phi - \gamma_2 \cos\phi) = \gamma^5
      \gamma_\mu g^\mu \label{eq:gp2}
    \end{equation}
    This gives Eqns.~(\ref{eq:spinsumzeta}) and (\ref{eq:spinsumxi}),
    the form
    \begin{subequations} \begin{eqnarray} && \sum_\alpha
	\xi_\alpha(\p) \stackrel{\neg}{\xi}_\alpha(\p) = m \left[
	  \gamma^5 \gamma_\mu g^\mu + \1 \right]
	\label{eq:spinsumxizimpok}\\ 
	&& \sum_\alpha \zeta_\alpha(\p)
	\stackrel{\neg}{\zeta}_\alpha(\p) = m \left[\gamma^5 \gamma_\mu
	  g^\mu -\1\right]\label{eq:spinsumzetazimpok} \end{eqnarray}
    \end{subequations}
    The $\gamma^\mu$, in the Weyl realisation, are taken to be
    \begin{equation}
      \gamma^0:=\left(\begin{array}{cc} \0 & \1 \\ \1 & \0
      \end{array}\right),\quad
      \gamma^i:=\left(\begin{array}{cc} \0 & -\sigma^i \\ \sigma^i & \0
      \end{array}\right),\quad 
      \gamma^5:= - i \gamma^0\gamma^1\gamma^2\gamma^3 =
      \left(\begin{array}{cc} \1 & \0 \\ \0 & -\1
      \end{array}\right)
    \end{equation}

    \section{Elko fermionic fields of mass dimension one: Lagrangian densities}
    \label{Sec:ElkoMDO}

    Confining to the Elko frame, we now examine the physical and
    mathematical content of two quantum fields\footnote{If one prefers,
      one may wish to consider $\Lambda(x)$ and $\lambda(x)$ as two
      mathematical objects; and not identify them as `quantum fields' in
      the sense of Weinberg's discourse in the opening volume of his
      triology on quantum fields.  Similar remarks apply to the use of
      terminology that we borrow from the well-known quantum fields.}

    \begin{equation}
      \Lambda(x) \stackrel{\rm def}{=} \int \frac{d^3 p}{(2\pi)^3}
      \frac{1}{\sqrt{2 m E(\p)}} \sum_\alpha{\Big[} a_\alpha(\p)
	\xi_\alpha(\p) e^{- i p_\mu x^\mu} +\; b^\ddagger_\alpha(\p)
	\zeta_\alpha(\p) e^{+ i p_\mu x^\mu} {\Big]}
    \end{equation}
    and,
    \begin{equation}
      \lambda(x)\stackrel{\rm def}{=}
      \Lambda(x)\big\vert_{b^\ddagger({\bf p}) \to a^\ddagger({\bf p})}
    \end{equation}
    We assume that the annihilation and creation operators satisfy the
    fermionic anticommutation relations
    \begin{subequations}
      \begin{eqnarray}
	\{a_\alpha(\p),\;a^\ddagger_{\alpha^\prime}(\p^\prime)\} = (2
	\pi)^3\, \delta^3(\p-\p^\prime)\,\delta_{\alpha\alpha^\prime},\\
	\{a_\alpha(\p),\;a_{\alpha^\prime}(\p^\prime)\} = 0,
	\quad\{a^\ddagger_\alpha(\p),\;a^\ddagger_{\alpha^\prime}
	(\p^\prime)\} =0.
      \end{eqnarray}
    \end{subequations}
    Similar anticommutators are assumed for the $b_\alpha(\p)$ and
    $b^\ddagger_\alpha(\p)$. The adjoint field
    $\stackrel{\neg}{\Lambda}(x)$ is defined as
    \begin{equation}
      \stackrel{\neg}\Lambda(x) \stackrel{\rm def}{=} \int \frac{d^3
	p}{(2\pi)^3} \frac{1}{\sqrt{2 m E(\p)}} \sum_\alpha{\Big[}
	a^\ddagger_\alpha(\p) \stackrel{\neg}{\xi}_\alpha(\p) e^{+ i p_\mu
	  x^\mu} +\; b_\alpha(\p) \stackrel{\neg}{\zeta}_\alpha(\p) e^{- i
	  p_\mu x^\mu} {\Big]}
    \end{equation}
    The results contained in Eqns.~(\ref{eq:a}-\ref{eq:d}) assure us
    that it is the Klein-Gordon, and not the Dirac, operator that
    annihilates the fields $\Lambda(x)$ and $\lambda(x)$. The
    associated Lagrangian densities are
    \begin{equation}
      {\mathcal L}^\Lambda(x) = \partial^\mu\stackrel{\neg}{\Lambda}(x)
      \partial_\mu\Lambda(x) - m^2
      \stackrel{\neg}{\Lambda}(x)\Lambda(x),\quad {\mathcal
	L}^\lambda(x) = {\mathcal L}^\Lambda(x)
      \big\vert_{\Lambda\rightarrow\lambda}\label{eq:Lagrangian}
    \end{equation}
    The mass dimensionality of these Elko fields is thus one, and not
    three half.

    \vspace{\baselineskip}

    The mass dimensionality of a field can also be deciphered from
    constructing the Feynman-Dyson propagator.  This matter is discussed
    in App.~\ref{App:TimeOrder}.

    \newpage

    \subsection{Identification of Elko with dark matter}
    \label{Sec:ElkoIdentificationWithDM}

    These results open up an entirely new and unexpected possibility for
    the dark matter sector. The primary observations that suggest this
    are five fold:

    \begin{enumerate}
    \item Due to mismatch in mass dimensionality of ${\mathcal
      D}_\Lambda=1$ and ${\mathcal D}_\lambda=1$ with the SM's matter
      fields ${\mathcal D}_\Psi=3/2$, the new fermionic fields cannot
      enter the SM doublets.

    \item The Lagrangian densities associated with Elko fields, do not
      carry the gauge symmetries of the SM (see remarks above
      Eqn.~(\ref{eq:gp2})).

    \item The dimension four interactions of the $\Lambda(x)$ and
      $\lambda(x)$ with the standard model fields are restricted to
      those with the SM Higgs doublet $\phi(x)$. These are
%      \begin{eqnarray}
%	&& {\mathcal L}^{\mathrm{int}}(x) =
%	\phi^\dagger(x)\phi(x)\,{\Big[} a_1\stackrel{\neg}{\Lambda}(x)
%	  {\Lambda}(x) + a_2 \stackrel{\neg}{\lambda}(x)%
%	  {\lambda}(x)\nonumber\\ &&\hspace{57pt}+ a_3 \left(
%	  \stackrel{\neg}{\Lambda}(x) {\lambda}(x) +
%	  \stackrel{\neg}{\lambda}(x) {\Lambda}(x)\right){\Big]}
 %    \end{eqnarray}
            \begin{equation}
	{\mathcal L}^{\mathrm{int}}(x) =
	\phi^\dagger(x)\phi(x)\,{\Big[} a_1\stackrel{\neg}{\Lambda}(x)
	  {\Lambda}(x) + a_2 \stackrel{\neg}{\lambda}(x)
	  {\lambda}(x)\nonumber\\ + a_3 \left(
	  \stackrel{\neg}{\Lambda}(x) {\lambda}(x) +
	  \stackrel{\neg}{\lambda}(x) {\Lambda}(x)\right){\Big]}
      \end{equation}
      where the $a's$ are unknown coupling constants and.
      
    \item
      By virtue of their mass dimensionality the new dark matter fields
      are endowed with dimension four self interactions
 %     \begin{eqnarray} 
%	&& {\mathcal L}^{\mathrm{self}} (x)= b_1
%	\left(\stackrel{\neg}{\Lambda}(x) {\Lambda}(x)\right)^2 + b_2
%	\left(\stackrel{\neg}{\lambda}(x) {\lambda}(x)\right)^2
%	\nonumber \\ &&\qquad + b_3 \left[\left(\stackrel{\neg}
%	  {\Lambda}(x) {\lambda}(x)\right)^2 +
%	  \left(\stackrel{\neg}{\lambda}(x) {\Lambda}(x)\right)^2\right],
   %   \end{eqnarray}
   \begin{equation} 
	 {\mathcal L}^{\mathrm{self}} (x)= b_1
	\left(\stackrel{\neg}{\Lambda}(x) {\Lambda}(x)\right)^2 + b_2
	\left(\stackrel{\neg}{\lambda}(x) {\lambda}(x)\right)^2
	+ b_3 \left[\left(\stackrel{\neg}
	  {\Lambda}(x) {\lambda}(x)\right)^2 +
	  \left(\stackrel{\neg}{\lambda}(x) {\Lambda}(x)\right)^2\right],
      \end{equation}
      where the $b's$ are unknown coupling constants. Observational
      evidence suggests that dark matter needs to be self
      interacting~\cite{Spergel:1999mh,Wandelt:2000ad,Ahn:2004xt,Balberg:2002ue}.\footnote{We
      parenthetically remark that, the interactions with the standard
      model gauge fields~\textendash~with
      $F^{\mathrm{SM}}_{\mu\nu}(x)$ symbolically representing the
      associated field strength tensors~\textendash~through Pauli
      terms
	\[
	  {\mathcal L}^{Pauli} (x)=
	  \stackrel{\neg}{\Lambda}(x)[\gamma^\mu,\gamma^\nu] {
	    \lambda}(x) F^{\mathrm{SM}}_{\mu\nu}(x),\quad \mbox{etc.}
	  \]
	  may in principle exist. However, we consider them to have
	  vanishing coupling strength as ${\mathcal L}^\Lambda(x)$ and
	  ${\mathcal L}^\lambda(x)$ do not carry invariance under SM
	  gauge transformations.  }

    \item The Elko fields are endowed with an intrinsic preferred
      axis. Tentative evidence already exists for such an
      axis~\cite{Land:2006bn,Samal:2008nv,Frommert:2009qw}.
      
    \end{enumerate}

    Combined, the enumerated Elko properties not only render Elko dark
    with respect to the SM matter fields but they also endow it
    various observationally-attractive properties. It is worth
    emphasising that all of these properties are intrinsic to Elko,
    and arise in a natural way.

    \subsection{The locality structure of Elko}

    The canonically conjugate momenta to the Elko fields are
    \begin{equation}
      \Pi(x)= \frac{\partial{\mathcal L}^\Lambda} {\partial\dot\Lambda}
      = \frac{\partial}{\partial t}\stackrel{\neg}{\Lambda}(x),\quad
    \end{equation}
    and similarly $ \pi(x) = \frac{\partial}{\partial
      t}\stackrel{\neg}{\lambda}(x)$. The calculational details for the
    two fields now differ significantly. We begin with the evaluation of
    the equal time anticommutator for the $\Lambda(x)$ and its conjugate
    momentum, and find
    \begin{equation} \{\Lambda(\x,t),\; \Pi(\x^\prime,t)\} =
      i\int\frac{d^3 p}{(2\pi)^3}\frac{1}{2 m} e^{i {\mathbf p}\cdot
	({\mathbf x}-{\mathbf x}^\prime)} \underbrace{\sum_\alpha\left[
	  \xi_\alpha(\p) \stackrel{\neg}{\xi}_\alpha(\p) - \zeta_\alpha(-
	  \p) \stackrel{\neg}{\zeta}_\alpha(- \p)\right]}_{=\,2 m [\1 +
	  {\mathcal G}(\mathbf{p})]}.
    \end{equation}
    or, equivalently
    \begin{equation}
      \{\Lambda(\x,t),\; \Pi(\x^\prime,t)\} = i \delta^3(\x -\x^\prime)
      \1 + i\int\frac{d^3 p}{(2\pi)^3} e^{i {\mathbf p}\cdot ({\mathbf
	  x}-{\mathbf x}^\prime)} {\mathcal G}(\mathbf{p}) .\label{eq:LPac}
    \end{equation}
    The anticommutators for the particle/antiparticle annihilation and
    creation operators suffice to yield the remaining locality
    conditions,
    \begin{equation}
      \{\Lambda(\x,t),\; \Lambda(\x^\prime,t)\} = \mathbb{O},\quad
      \{\Pi(\x,t),\; \Pi(\x^\prime,t)\}
      \label{eq:LLPPac} = \mathbb{O}.
    \end{equation}
    Since the integral on the right hand side of Eqn.~(\ref{eq:LPac})
    vanishes only along the $\pm \hat z_e$ axis, the preferred axis also
    becomes the \emph{axis of locality.}

    For the equal time anticommutator of the $\lambda(x)$ field with its
    conjugate momentum, we find
    \begin{equation}
      \{\lambda(\x,t),\; \pi(\x^\prime,t)\} = i\int\frac{d^3
	p}{(2\pi)^3}\frac{1}{2 m} \sum_\alpha\left[ e^{i {\mathbf p}\cdot
	  ({\mathbf x}-{\mathbf x}^\prime)} \left(\xi_\alpha(\p)
	\stackrel{\neg}{\xi}_\alpha(\p) - \zeta_\alpha(- \p)
	\stackrel{\neg}{\zeta}_\alpha(- \p)\right)\right].
    \end{equation}
    Which, using the same argument as before, yields
    \begin{equation}
      \{\lambda(\x,t),\; \pi(x^\prime,t)\} = i \delta^3(\x -\x^\prime)
      \1 + i\int\frac{d^3 p}{(2\pi)^3} e^{i {\mathbf p}\cdot ({\mathbf
	  x}-{\mathbf x}^\prime)} {\mathcal G}(\mathbf{p}).\label{eq:lpac2}
    \end{equation}
    The difference arises in the evaluation of the remaining
    anticommutators.  The equal time $\lambda$-$\lambda$ anticommutator
    reduces to
    \begin{equation} 
      \{\lambda(\x,t),\; \lambda(\x^\prime,t)\} = \int\frac{d^3
	p}{(2\pi)^3}\frac{1}{2 m E(\p)}\; e^{i {\mathbf p}\cdot ({\mathbf
	  x}-{\mathbf x}^\prime)} \underbrace{\sum_\alpha\left[
	  \xi_\alpha(\p) \zeta^T_\alpha(\p) + \zeta_\alpha(- \p)
	  \xi^T_\alpha(- \p)\right]}_{:=\,\Omega(\p)}.  \label{eq:zimpokb}
    \end{equation}
    Now using explicit expressions for $ \xi_\alpha(\p)$ and $
    \zeta_\alpha(\p)$ we find that $\Omega(\p)$ identically vanishes.
    Eqn.~(\ref{eq:zimpokb}) then implies
    \begin{equation}
      \{\lambda(\x,t),\; \lambda(\x^\prime,t)\} = \mathbb{O}.
      \label{eq:llac2}
    \end{equation}
    Finally, the equal time $\pi$-$\pi$ anticommutator simplifies to
%    \begin{eqnarray}
 %    &&\{\pi(\x,t),\; \pi(\x^\prime,t)\} = \int\frac{d^3
%	p}{(2\pi)^3}\frac{E(\p)}{2 m}\; e^{-i {\mathbf p}\cdot ({\mathbf
%	  x}-{\mathbf x}^\prime)} \nonumber\\
%      &&\times\underbrace{\sum_\alpha\left[
%	  \Big(\stackrel{\neg}{\xi}_\alpha(\p) \Big)^T
%	  \stackrel{\neg}{\zeta}_\alpha(\p) +
%	  \Big(\stackrel{\neg}{\zeta}_\alpha(-\p) \Big)^T
%	  \stackrel{\neg}{\xi}_\alpha(-\p) \right]}_{=\mathbb{O},~
%	\mbox{by a direct evaluation }},\nonumber
%    \end{eqnarray}
  \begin{equation}
      \{\pi(\x,t),\; \pi(\x^\prime,t)\} = \int\frac{d^3
	p}{(2\pi)^3}\frac{E(\p)}{2 m}\; e^{-i {\mathbf p}\cdot ({\mathbf
	  x}-{\mathbf x}^\prime)} 
      \underbrace{\sum_\alpha\left[
	  \Big(\stackrel{\neg}{\xi}_\alpha(\p) \Big)^T
	  \stackrel{\neg}{\zeta}_\alpha(\p) +
	  \Big(\stackrel{\neg}{\zeta}_\alpha(-\p) \Big)^T
	  \stackrel{\neg}{\xi}_\alpha(-\p) \right]}_{=\mathbb{O},~
	\mbox{by a direct evaluation }},\nonumber
    \end{equation}
    yielding
    \begin{equation}
      \{\pi(\x,t),\; \pi(\x^\prime,t)\} = \mathbb{O}. \label{eq:ppac2}
    \end{equation}
    Equations (\ref{eq:LPac}-\ref{eq:LLPPac}) and
    (\ref{eq:lpac2}-\ref{eq:ppac2}) establish that $\Lambda(x)$ and
    $\lambda(x)$ are {\em local} quantum fields in the direction
    perpendicular to the Elko plane; i.e. along the preferred axis $\hat
    z_e$.  We propose to call $\hat{z}_e$ as the \emph{axis of locality}
    in the dark sector.

    \section{Concluding remarks}
    \label{Sec:ConcludingRemarks}
    Modulo its specilisation to the Majorana field, Weinberg's
    monographic work~\cite{Weinberg:1995mt} establishes the uniqueness
    of the Dirac quantum field for spin one half particles. Seen from
    that perspective the Ahluwalia-Grumiller work on Elko in 2005 was
    unexpected. At this stage two things happened. On the one hand
    Elko found significant interest among mathematical physicists and
    cosmologists~\cite{Boehmer:2007dh,Boehmer:2006qq,daRocha:2005ti,HoffdaSilva:2009is,daRocha:2008we,daRocha:2007pz,Boehmer:2009aw,Shankaranarayanan:2009sz,Boehmer:2008ah,Gredat:2008qf,Boehmer:2008rz,Boehmer:2007ut,daRocha:2007sd,Fabbri:2009ka}.
    In these papers one dealt with Elko as spinors and not as a
    quantum field. So no contradiction with Weinberg's theorem-like
    work occurred.  Not unexpectedly, Gillard and Martin showed that
    if Elko were to be taken as `good' quantum fields, Poincar\'e
    symmetries had to be violated in some form or
    other~\cite{Gillard:2009zw}. The results presented in this
    communication explicitly confirm this and show that the violation
    occurs in a rather subtle way. This done, Elko now stands as a
    natural dark matter candidate.  By virtue of its mass
    dimensionality it allows an unsuppressed quartic self
    coupling. Additionally, it points towards the existence of a
    preferred axis, along which the Elko quantum fields become local.
    Both of these aspects can be used to distinguish it from other
    candidates for the dark-matter sector. Its darkness with respect
    to the SM matter and gauge fields is built into its intrinsic mass
    dimension.  \acknowledgments

    We thank Adam Gillard, Ben Martin, and Thomas Watson for their
    constant questions and discussions, and also Karl-Henning Rehren for
    his helpful comments.  We are also grateful to a JCAP referee for a
    careful reading of the manuscript and for sharing his/her
    questions with us. The presentation of our draft was improved because of a
    discussion with Matt Visser.

    \appendix
    \section{Appendix}

    \subsection{Dirac spin sums and  a `misleading' derivation of Dirac equation}
    \label{App:MisleadingDerivation}

    With a minor departure from the historical path, the Dirac
    counterpart of (\ref{eq:spinsumxizimpok}) and
    (\ref{eq:spinsumzetazimpok}) may be constructed as follows.  Instead
    of (\ref{eq:MajoranaSpinor}), we start with
    \begin{equation} \psi_D \stackrel{\rm def}{=} \left(\begin{array}{cc} \phi_r 
	\\ \phi_\ell\end{array}\right)
	\label{eq:DiracSpinor}
    \end{equation} 
    The helicities of $\phi_r$ and $\phi_\ell$ are identical and are
    determined by requiring that $\psi_D$ be eigenspinors of the
    parity operator $S(P)$.  Again, there are four independent rest
    spinors (these differ from those mentioned in Sec.~(\ref{Sec:SM})
    only in that we now work in the `helicity basis')
    \begin{eqnarray}
      u_{+1/2}(\ep) &=& \left(\begin{array}{r} \phi_r^+(\ep)\\
	\phi_\ell^+(\ep)\end{array}\right),
      \hspace{.68cm} u_{-1/2}(\ep) = \left(\begin{array}{r} \phi_r^-(\ep) \\
	\phi_\ell^-(\ep)\end{array}\right) \label{eq:DiracSpinoru}\\
      v_{+1/2}(\ep) &=& \left(\begin{array}{r}
	\phi_r^-(\ep)\\-\phi_\ell^-(\ep)\end{array}\right),
      \hspace{.4cm} v_{-1/2}(\ep) = \left(\begin{array}{r} -\phi_r^+(\ep) \\ \phi_\ell^+(\ep)\end{array}\right) \label{eq:DiracSpinorv}
    \end{eqnarray} 
    The $u(\p)$ and $v(\p)$ for an arbitrary momentum are obtained via
    the action of the boost $\kappa$
    \begin{equation}
      u(\p) = \kappa\, u(\ep),\quad v(\p) = \kappa\, v(\ep)
    \end{equation}
    These lead to the spin sums
    \begin{subequations} 
      \begin{eqnarray} && \sum_\beta
	u_\beta(\p) \overline{u}_\beta(\p) = m \left[ \frac{\gamma_\mu
	    p^\mu}{m} + \1 \right]
	\label{eq:spinsumu}\\ 
	&& \sum_\beta v_\beta(\p) \overline{v}_\beta(\p) = m \left[
	  \frac{\gamma_\mu p^\mu}{m} -\1\right]\label{eq:spinsumv}
      \end{eqnarray}
    \end{subequations}
    where $\beta$ takes two values: $+1/2$ and $-1/2$. As before, the
    right hand sides in the above expression simply express the result
    of a direct evaluation of the left hand sides.  These are covariant.

    We thus see that in the Dirac construct, whether it be at the level
    of spinors (or at the quantum field theoretic level), no preferred
    frame is introduced.  For Majorana spinors, and Elko, the conclusion
    is both unexpected and inevitable. This difference \textemdash~as
    pertaining to the existence of a preferred frame \textemdash~between
    the Dirac and Majorana spinors, along with their cousins Elko, to
    our knowledge is completely unknown. This conclusion carries
    distinct echoes of the unpublished notes~\cite{Ahluwalia:2003jt}
    which eventually, in collaboration with Grumiller, led to the
    discovery reported in
    references~\cite{Ahluwalia:2004sz,Ahluwalia:2004ab}.

    If we multiply Eqn.~(\ref{eq:spinsumu}) by $u_{\beta^\prime}(\p)$
    from the right, and use $\overline{u}_\beta(\p) u_\beta^\prime(\p)
    = 2 m \delta_{\beta\beta^\prime}$; and carry out a similar
    exercise with Eqn.~(\ref{eq:spinsumu}), then after a minor
    rearranging we obtain
    \begin{eqnarray}
      && \left(\gamma_\mu p^\mu - m \1\right) u(\p) = 0 \\ &&
      \left(\gamma_\mu p^\mu + m \1\right) v(\p) = 0
    \end{eqnarray}
    These are indeed Dirac equations in momentum space. With $\p^\mu
    \to i\partial^\mu$ and
    \begin{equation}
      \psi(x) \stackrel{\rm def}{=}\bigg\{\begin{array}{l} u(\p)\exp(
      - i p_\mu x^\mu)\\ v(\p)\exp( + i p_\mu x^\mu)
      \end{array}     
    \end{equation}
    these yield the well-known Dirac equation in the configuration space
    \begin{equation}
      \left(i \gamma_\mu\partial^\mu -m \1\right)\psi(x) = 0
    \end{equation}
    To associate these with the dynamics of spin one half spinors,
    particularly in a quantum field theoretic context (where $\psi(x)$
    is now elevated to a spinor field $\Psi(x)$), requires that, in
    addition, the vacuum expectation value, $\langle\;\vert{\mathcal
    T}[\Psi(x^\prime)\,\overline{\Psi}(x)] \vert\;\rangle$, be
    proportional to the relevant Green's function. That is, it is not
    sufficient to find an operator, such as $\left(i
    \gamma_\mu\partial^\mu -m \1\right)$, or the Klein Gordon
    operator, that annihilates $\Psi(x)$ for it to serve in the
    Lagrangian density of the field $\Psi(x)$. It must also satisfy
    the said requirement. This will become abundantly clear from what
    follows in the context of Elko.

    While we do consider the above `derivation' of Dirac equation
    misleading, it does serve to tell us that the Dirac spinors are
    eigenspinors of $\gamma_\mu p_\mu$ with eigenvalues $\pm m$.
    \begin{equation}
      \gamma_\mu p^\mu u(\p) = + m u(\p),\quad \gamma_\mu p^\mu v(\p)
      = - m v(\p)
    \end{equation}
    The Elko counterpart is
    \begin{equation}
      \mathcal{G}(\p)\xi(\p) = + \xi(\p),\quad 
      \mathcal{G}(\p)\zeta(\p) = - \zeta(\p)\label{eq:Gq}
    \end{equation}
    It again emphasises that identities such as these should not be
    mistaken for dynamical equations. In particular,
    $\mathcal{G}(\p)$, unlike its Dirac counterpart $ \gamma_\mu
    p^\mu$, contains no time derivative.

    \subsection{Elko time ordering and propagators}
    \label{App:TimeOrder}

    The mass dimensionality of a field can also be deciphered from
    constructing the Feynman-Dyson propagator. This involves defining
    a time ordering operator. However, the existence of a preferred
    direction mentioned above makes it unclear as to how, and if, Elko
    construct modifies this definition. In what follows we first adopt
    the standard definition of the fermionic time ordering operator,
    and then invoke a consistency argument to re-define the Elko time
    ordering.
    
    Let $\mathcal T$ be the standard fermionic time ordering
    operator. Then, a straightforward calculation yields
    \begin{eqnarray}
      \hspace{-21pt}\langle\;\vert{\mathcal T}
             [\Lambda(x^\prime)\,\stackrel{\neg}{\Lambda}(x)]
             \vert\;\rangle && = \int \frac{d^3 p}{(2\pi)^3}\frac{1}{2
               m E(\p)} \times \sum_\alpha{\Big[} \theta(t^\prime - t)
               \xi_\alpha(\p) \stackrel{\neg}{\xi}_\alpha(\p) e^{-i
		 p_\mu(x^{\prime\mu} - x^\mu)} \nonumber\\ &&\qquad -\,
               \theta(t - t^\prime)\zeta_\alpha(\p)
               \stackrel{\neg}{\zeta}_\alpha(\p) e^{+i
		 p_\mu(x^{\prime\mu} - x^\mu)}{\Big]}\label{eq:vexpv}
    \end{eqnarray}
    where the step function $\theta(t)$ equals unity for $t>0$ and
    vanishes for $t<0$. 

    Using the spin sums (\ref{eq:spinsumxi}) and
    (\ref{eq:spinsumzeta}), setting $q^\mu=(q^0,{\mathbf q}=\p)$, and
    using the standard integral representation for the $\theta(t)$,
    Eqn.~(\ref{eq:vexpv}) simplifies to
    \begin{equation}
      \hspace{-21pt}\langle\;\vert{\mathcal T}
             [\Lambda(x^\prime)\,\stackrel{\neg}{\Lambda}(x)]
             \vert\;\rangle = i \int \frac{d^4 q}{(2\pi)^4} e^{-i
               q_\mu(x^{\prime\mu}- x^\mu)} \left[ \frac{\1 + {\mathcal
		   G}(\q)} {q_\mu q^\mu - m^2 + i\epsilon}\right]
	     \label{eq:zimpok}
    \end{equation}
    where the limit $\epsilon\to 0^+$ is understood.\footnote{The
      substitution through $q^\mu$ requires some discussion; see
      Sec.~6.2 of Ref.~\cite{Weinberg:1995mt} for details.} If there
    was no preferred axis then the integral involving the $
    {\mathcal G}(\q)$ term would have identically
    vanished. Consistency with result~(\ref{eq:Lagrangian}) suggests
    that in Elko quantum field theory one may need to modify the
    definition of the time ordered product to ${\mathcal{T}}_\#$,
    such that
    \begin{equation}
      \hspace{-21pt}\langle\;\vert{\mathcal {T}}_\#
             [\Lambda(x^\prime)\,\stackrel{\neg}{\Lambda}(x)]
             \vert\;\rangle = i \int \frac{d^4 q}{(2\pi)^4} e^{-i
               q_\mu(x^{\prime\mu}- x^\mu)} \left[ \frac{\1 } {q_\mu
		 q^\mu - m^2 + i\epsilon}\right]
	     \label{eq:zimpokzzz}
    \end{equation}
    To decipher the mass dimensionality, let ${\mathcal D}_\Lambda$ be
    the mass dimensionality of $\Lambda(x)$. Then the left-hand side
    of the above equation has mass dimension $2 {\mathcal
    D}_\Lambda$. As for the right hand side, the mass dimensionality
    is $2$. This gives ${\mathcal D}_\Lambda=1$.  Similarly, a simple
    computation shows that $\langle\;\vert{\mathcal T}_\#
    [\Lambda(x^\prime)\,\stackrel{\neg}{\Lambda}(x)] \vert\;\rangle =
    \langle\;\vert{\mathcal T}_\#
    [\lambda(x^\prime)\,\stackrel{\neg}{\lambda}(x)] \vert\;\rangle
    $. As such, ${\mathcal D}_\lambda = 1$.

  Applying the operator
  $\left[\partial^{\prime\mu}\partial^{\prime}_\mu + m^2\right]$ from
  the left on both sides of Eqn.~(\ref{eq:zimpokzzz}) gives \beq
  \left[\partial^{\prime\mu} \partial^{\prime}_\mu + m^2\right]
  \langle\;\vert{\mathcal T}_\#
  [\Lambda(x^\prime)\,\stackrel{\neg}{\Lambda}(x)] \vert\;\rangle = -i
  \delta^4(x^{\prime\mu} -x^\mu) \eeq 

\bls
In comparison, for the Dirac field
\begin{equation}
\hspace{-21pt}\langle\;\vert{\mathcal
T}[\Psi(x^\prime)\,\overline{\Psi}(x)]\vert\;\rangle = i \int\frac{d^4
q}{(2\pi)^4} e^{-i q_\mu(x^{\prime\mu}- x^\mu)}\left[ \frac{m \1 +
{\gamma^\mu q_\mu}} {q_\mu q^\mu - m^2 +
i\epsilon}\right]\label{eq:zimpokDirac}
\end{equation}
This well-known result gives, ${\mathcal D}_\Psi=\frac{3}{2}$. The
reader is reminded that the ${\gamma^\mu q_\mu}$ structure appears
here through the spins sums which, in the logical framework of this
communication, do not invoke any wave equation or a Lagrangian
density. Applying the operator $\left[i\gamma^\mu\partial^{\prime}_\mu
- m\right]$ from the left on both sides of Eqn.~(\ref{eq:zimpokDirac})
yields \beq \left[i\gamma^\mu\partial^\prime_\mu - m\right]
\langle\;\vert{\mathcal T} [\Psi(x^\prime)\,\overline{\Psi}(x)]
\vert\;\rangle = i\delta^4(x^{\prime\mu} -x^\mu) \eeq

\bibliography{DM2009PRD}

\end{document}